\documentclass[conference]{IEEEtran}
\IEEEoverridecommandlockouts
\usepackage{cite}
\usepackage{amsmath,amssymb,amsfonts}
\usepackage{algorithmic}
\usepackage{graphicx}
\usepackage{textcomp}
\usepackage{xcolor}
\usepackage{booktabs}
\usepackage{hyperref}
\def\BibTeX{{\rm B\kern-.05em{\sc i\kern-.025em b}\kern-.08em
    T\kern-.1667em\lower.7ex\hbox{E}\kern-.125emX}}

\IEEEpubid{\\\\ \begin{minipage}{\textwidth}\ \\[8pt]
  979-8-3315-3183-6/25/\$31.00 \copyright 2025 IEEE \\
\end{minipage}}

\begin{document}

\title{Enhancing Non-Intrusive Load Monitoring with Features Extracted by Independent Component Analysis\\
}

\author{

\IEEEauthorblockN{Sahar Moghimian Hoosh$^{1,2,a}$, Ilia Kamyshev$^{1,2,b}$, Henni Ouerdane $^{1,c}$}
\IEEEauthorblockA{$^{1}$ Center for Digital Engineering, Skolkovo Institute of Science and Technology, Moscow, Russian Federation \\
$^{2}$ Monisensa Development LLC, Moscow, Russian Federation\\
$^{a}$sahar.moghimian@skoltech.ru, $^{b}$ilia.kamyshev@skoltech.ru} $^{c}${h.ouerdane@skoltech.ru} 


}
\maketitle

\begin{abstract}

In this paper, a novel neural network architecture is proposed to address the challenges in energy disaggregation algorithms. These challenges include the limited availability of data and the complexity of disaggregating a large number of appliances operating simultaneously. The proposed model utilizes independent component analysis as the backbone of the neural network and is evaluated using the F1-score for varying numbers of appliances working concurrently. Our results demonstrate that the model is less prone to overfitting, exhibits low complexity, and effectively decomposes signals with many individual components. Furthermore, we show that the proposed model outperforms existing algorithms when applied to real-world data.

\end{abstract}


\section{Introduction}
\label{submission}

Non-intrusive load monitoring, also known as energy disaggregation, is the technique of breaking down a household's total energy use into individual appliance-level components using advanced analysis\cite{angelis2022nilm}. This technique has received significant attention in recent years due to its potential to enable greater energy efficiency, demand response, and load forecasting \cite{Event_detection}. Energy disaggregation has been extensively studied in the literature, and a variety of techniques have been proposed to perform this task \cite{faustine2017survey}
The concept of NILM was first introduced in the 1980s by G. Hart \cite{hart1992nonintrusive}, and since then, it has been a popular topic of research in the field of energy management. 

Various techniques have been suggested to enhance the accuracy of NILM, which, however, can be affected by several factors such as the total number of appliances, the number of appliances that are working simultaneously, the types of appliances, and the measurement noise. Multiple simultaneous appliance switching detection, and correct estimation in practical scenarios with noisy data, remain to be addressed robustly. Recent studies in deep learning models for NILM \cite{en13092195}, indicate that training deep neural networks on the limited labeled data can lead to reduced disaggregation accuracy, increased generalization error, and overfitting \cite{angelis2022nilm}. Thus, algorithm complexity and a lack of datasets continue to be major obstacles in the field of energy disaggregation \cite{Rafiq2021GeneralizabilityIO}. 

The typical algorithms are heavy in memory and sensitive to overfitting, which makes them difficult to be ported to the sensor. Moreover, most of the algorithms are trained on a limited number of appliances, while datasets contain dozens of classes of appliances on average. There is also a lack of related studies on the ``goodness’’ of disaggregation, with numerous individual components presented in the aggregated signal. The available datasets have limited combinations of different appliances, biased towards the most frequently used ones \cite{kamyshev2021cold}. These challenges need to be addressed to improve the accuracy and efficiency of energy disaggregation algorithms.

This paper presents a study on the performance of energy disaggregation algorithms for a varying number of individual appliances operating simultaneously. Additionally, we propose a novel neural network architecture that considers the physics of the energy disaggregation problem by utilizing an unmixing matrix obtained through independent component analysis (ICA).  
In this work, we demonstrate that the proposed model is less prone to overfitting, exhibits low complexity, and can effectively decompose signals with a large number of individual components. The algorithm is evaluated by analyzing the impact of an increasing number of components on the algorithm’s classification accuracy. To the best of our knowledge, this research is the first to use ICA as a feature extraction technique to enhance the performance of the multi-label classification models in NILM, specifically with high-frequency sampling data ($>1 kHz$). The results of this study provide the valuable insights for the development of accurate and efficient energy disaggregation algorithms capable of handling complex scenarios and diverse datasets. This work is implemented in Python and is available via the provided \href{https://github.com/arx7ti/ML2023SK-final-project}{link}. 

The remaining of this paper is organized as follows: Section II reviews related works, focusing on existing methods and their limitations. Section III describes the real, and synthetic datasets we used for the study in detail. Section IV presents the proposed multi-label classification model and baseline algorithms. Section V outlines the experimental setup, followed by Section VI, which presents the results and analysis. Finally, in Section VII, we summarize major findings and offer suggestions for further study.

\section{Related Works}\label{sec:related_works}

In recent years, efforts have been invested to adapt machine learning and deep learning techniques to the NILM domain due to significant advancements in hardware computing power and the improved performance of these techniques in various pattern recognition problems \cite{gopinath2020energy}. For instance, in \cite{cunado2019supervised}, the authors utilized the k-NN algorithm with steady-state features to identify individual appliances, achieving good performance. However, the approach was not very effective for detecting unknown appliances. The proposed approach by \cite{held2018frequency} offers a significant advantage as the FIT-PS signal representation fully retains the information of the current signal, including the phase shift. The paper suggests using this representation as a feature for a feedforward neural network, which leads to more robust results compared to standard approaches that use active power, reactive power, and a fixed number of harmonics. The superiority of the FIT-PS representation is particularly evident in scenarios with numerous simultaneous appliance activations. Additionally, the use of Long short-term memory (LSTM) nets with FIT-PS representation further improves the classification accuracy by incorporating transient state features. However, determining the best initialization for LSTM nets necessitates the use of a validation set due to the problem of multiple local minima. The approach based on deep convolutional neural networks using techniques from image segmentation was applied for the NILM problem in paper \cite{massidda2020non}. Our choice of deep convolutional layers is motivated by its automatic feature extraction capabilities as raw load active power signals are given as input. The core part of the proposed architecture is the temporal pooling module that expands the feature dimensionality of input for multilabel classification of multiple active loads operating simultaneously. However, the expansion of the feature space in the encoder and pooling layers are done at a cost of time resolution reduction of the input signal.

\section{Datasets}\label{sec:dataset}
As we are focusing on signals with a high sampling rate to track the electrical properties of an appliance, we selected the PLAID dataset \cite{gao2014plaid}, which comprises 1800 samples measured at 30 kHz, with a total number of classes equal to 16. Each sample is a pair of voltage and current signals related to one particular class of appliances. In addition to the real data, we generated synthetic linearly separable classes of appliances based on the method presented in \cite{kamyshev2023physics}. For this purpose, we used given sub-metered measurements (signals of individual appliances) and artificially composed them into aggregated signals. This procedure is additionally fair from a physical point of view as long as all the measurements are from the same power grid, i.e., the same voltage level, frequency, etc. Likely, this condition is satisfied. In real circuits, the same process obeys the Kirchhoff law. 

A recent study \cite{kamyshev2021cold} showed that the most frequent number of simultaneously working appliances is equal to eight for a typical household. Since the goal of this research is to investigate the model’s performance on numerous components, we generated samples of aggregated signals where from one to $n_{\mathrm{classes}}$ types of appliances appear at the same time. Moreover, we also accounted for the fact that some appliances may be duplicated from one to ten times. This is because one particular type of appliance in a household may be presented multiple times, e.g., three phone chargers, two air-conditioners, ten light bulbs. Aggregated samples were generated in a random manner, where each class has an equal probability of being presented in the mixture. By doing so, we aimed to reduce the class imbalance of the resulting dataset.

To prepare dataset, first, we resampled signals of the PLAID dataset to 3 kHz. Then we extracted 19,000 regions of interest from the PLAID dataset and generated the same amount of synthetic regions of interest related to standalone appliances. We split them into subsets in the following proportions 70\%, 10\% and 20\%, for train, validation and test respectively. Further, each subset was used to generate accordingly 7000, 1000, and 2000 of aggregated samples. Each aggregated sample is represented by a binary vector, where a value of 1 indicates that the corresponding appliance class is included in the mixture.  

\section{Algorithms and Models}\label{sec:algorithms}

\subsection{Proposed model: ICA+ResNetFFN}
\begin{figure*}
\vskip 0.2in
\begin{center}
\centering
\centerline{\includegraphics[width=1.5\columnwidth]{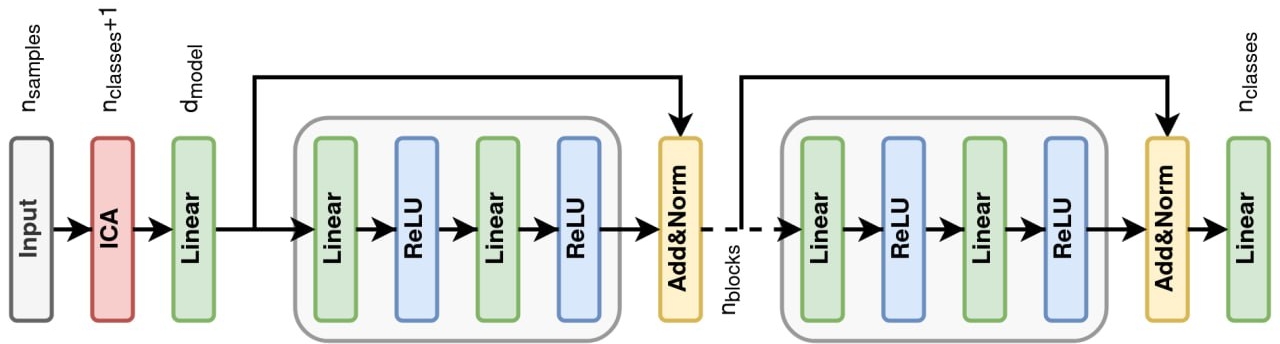}}
\caption{Architecture of the proposed model, ICA+ResNetFFN. The number of parameters for this model is 65,000.}
\label{proposed_model}
\end{center}
\vskip -0.2in
\end{figure*}

The proposed neural network architecture is shown on Figure \ref{proposed_model}. 
The incoming aggregated signal \( X \) is decomposed into \( n_{\mathrm{classes}} + 1 \) components using the un-mixing matrix \( U \), obtained by Independent Component Analysis (FastICA realization), i.e., $X’=XU^T$. The term “+1” accounts for a Gaussian component. ICA is used because it assumes that sources are mixed linearly. This assumption aligns with the physics of the power grids. Specifically, the aggregated signal measured at the power grid's input node adheres to Kirchhoff's law, $i_{agg}(t)=\sum_k{i_k(t)}$ where \( i_k(t) \) represents the individual currents from each source \( k \). Thus, the aggregated signal is a linear mixture of these individual source contributions. Once $X’$ is obtained, it is linearly projected to a space of dimension $d_{\mathrm{model}}$, i.e. $X_d=X’W^T+b=XU^TW^T+b$. Furthermore, $X_d$ is passed through the sequence of $n_{\mathrm{blocks}}$ paired linear layers, followed by ReLU activations and residual connections, as shown in Figure \ref{proposed_model}. We used $d_{\mathrm{model}}=64$ and $n_{\mathrm{blocks}}=15$ in this project. By implementing this relatively simple architecture, we aim to show that the understanding of the process and nature of the data may guide the selection of an algorithm.

\subsection{Baseline models}

\subsubsection{Temporal Pooling NILM}

The Temporal Pooling NILM (TP-NILM) architecture, adopted from reference \cite{massidda2020non}, is utilized for decomposing the aggregated signal and identifying appliance activation states. This network comprises encoder, temporal pooling, and decoder modules. The encoder module transforms the one-dimensional discrete input signal into a higher-dimensional feature space, consisting of a series of convolutional and max pooling layers with ReLU activation, followed by batch normalization and a dropout layer for regularization. While the encoder reduces the input signal's time resolution, it extracts 256 output features from a single aggregated time-domain signal.

The temporal pooling module accepts the encoder output and generates additional temporal context-aware features using four average pooling layers with different filter settings. These pooling layers further decrease the signal's time resolution while maintaining the feature dimension consistent with the encoder output. The four signals then pass through a convolutional layer with ReLU activation and batch normalization. Additionally, the temporal pooling module performs upsampling to approximate the encoder output's time resolution, merging the detailed features from the encoder module with the additional temporal context features from the temporal pooling module, and passing them to the decoder module. The decoder module reduces the input's feature dimension by passing it through a convolutional layer. Subsequently, the output of the convolutional layer is fed into a fully connected layer that returns an array of scores for appliance-specific activations. Finally, the sigmoid function is applied to obtain values within the range of (0, 1), and appliance activation states are determined using a decision threshold of 0.5.

\subsubsection{FIT-PS+LSTM}
The FIT-PS method is a novel signal processing method, which was successfully applied as a feature extraction method for classification in NILM \cite{held2018frequency}. It consists of three steps. The first step is to divide the sampled signal with respect to the fundamental frequency of the power grid. The division is done by finding the abscissa crossing, namely, the change from negative values of voltage to positive values. Abscissa crossing is chosen since it represents the point of maximum steepness, has an almost constant derivative in sinusoidal signals, and is less affected by amplitude variations compared to other parts of the signal. In the second step, abscissa crossing is used to estimate the linear approximation of absolute position of zero crossing. Next, linear interpolation is applied to assign indices to each individual period of the signal. The resulting matrix $X_{l,k}$ has $n_l \times n_k$ dimensions, where $n_l$ is the number of periods and $n_k$ is the number of sampling points in each period.

The signal passed through the FIT-PS method can be fed to the LTSM network, where $n_k$ is the input dimension.

An output of such a network is being averaged across a number of periods and then being passed through a fully connected layer with output size equal to $n_{\mathrm{classes}}$. Finally, a sigmoid layer is used to calculate scores of each class being present in the aggregated signal.  

\subsubsection{Fryze+CNN}
The CNN model proposed in \cite{faustine2020multi} uses the Fryze power theory \cite{staudt2008fryze} and the Euclidean distance matrix as feature extraction step for the multi-label classifier. Within the theory, the activation current is decomposed into orthogonal components related to electrical energy in the time-domain:
\begin{equation}
i(t)=i(t)_a+i(t)_f
\end{equation}
The active current $i(t)_{a}$ is the current passing through the resistive load. In Fryze’s theory, the active power is calculated as the average value of $i(t) \cdot v(t)$ over one fundamental cycle $T_{s}$ defined as follows;

\begin{equation}
i(t)_a=\frac{p_a}{v_{r m s}^2} v(t)
\end{equation}

where the rms voltage $v_{rms}$ is expressed as follows
\begin{equation}
v_{r m s}=\sqrt{\frac{1}{T_s} \sum_{t=1}^{T_s} v(t)^2}
\end{equation}

The non-active component is then equal to
\begin{equation}
i(t)_f=i(t)-i(t)_a
\end{equation}
The orthogonal components of the current, namely $i(t)_a$, and $i(t)_f$ undergo two pre-processing steps. Firstly, the signals are dimensionally reduced using the piece-wise aggregate approximation. Secondly, the distance matrix for each signal is computed. The two resulting distance matrices are then combined to create input for the multi-label classifier. The classifier is a four-block convolutional neural network, featuring 16, 32, 64 and 128 channels, with kernel sizes of $5 \times 5$, $5 \times 5$, $3 \times 3$, and $3 \times 3$ and strides of size 2, respectively. The remaining part of the network comprises three linear layers, consisting of 512, 1024 and $n_{\mathrm{classes}}$ neurons, respectively. The main activation function employed is ReLU.

\section{Experiments}\label{sec:experiments}
We conducted two experiments for the synthetic and real classes of appliances, respectively. During each experiment, we trained and validated six models: ICA+ResNetFFN (our proposed model), Fryze+CNN \cite{faustine2020multi}, Temporal Pooling NILM \cite{massidda2020non}, FIT-PS+LSTM \cite{held2018frequency}, ICA+k-NN, and ICA+Random Forest. The last two models were chosen as they showed significant performance on the ICA features \cite{giri2013towards}. Each experiment required 45 minutes of processing time on a machine with $2 \times$ RTX 2080 Ti GPUs and 128GB of RAM. 

We assessed the performance of classification algorithms using the $F_1$-score, averaged over samples. To get a complete picture of the disaggregation performance, we computed the $F_1$-score for each number of appliances present simultaneously, from 1 to $n_{\mathrm{classes}}$ and for the whole sample. Ideally, the distribution of $F_1$-scores across a different number of simultaneously working appliances should be uniform. That is, there should not be a bias towards a particular number of appliances at a runtime. Intuitively, with the number of appliances growing, the $F_1$-score is expected to drop. This is mainly due to the fact that low-power appliances such as charges, are treated as noise compared to the high-power appliances while they are operating at the same time.

\section{Results}\label{sec:results}
\subsection{Experiment 1: Synthetic Appliances}

\begin{figure}
    \centering
    \includegraphics[width=\columnwidth]{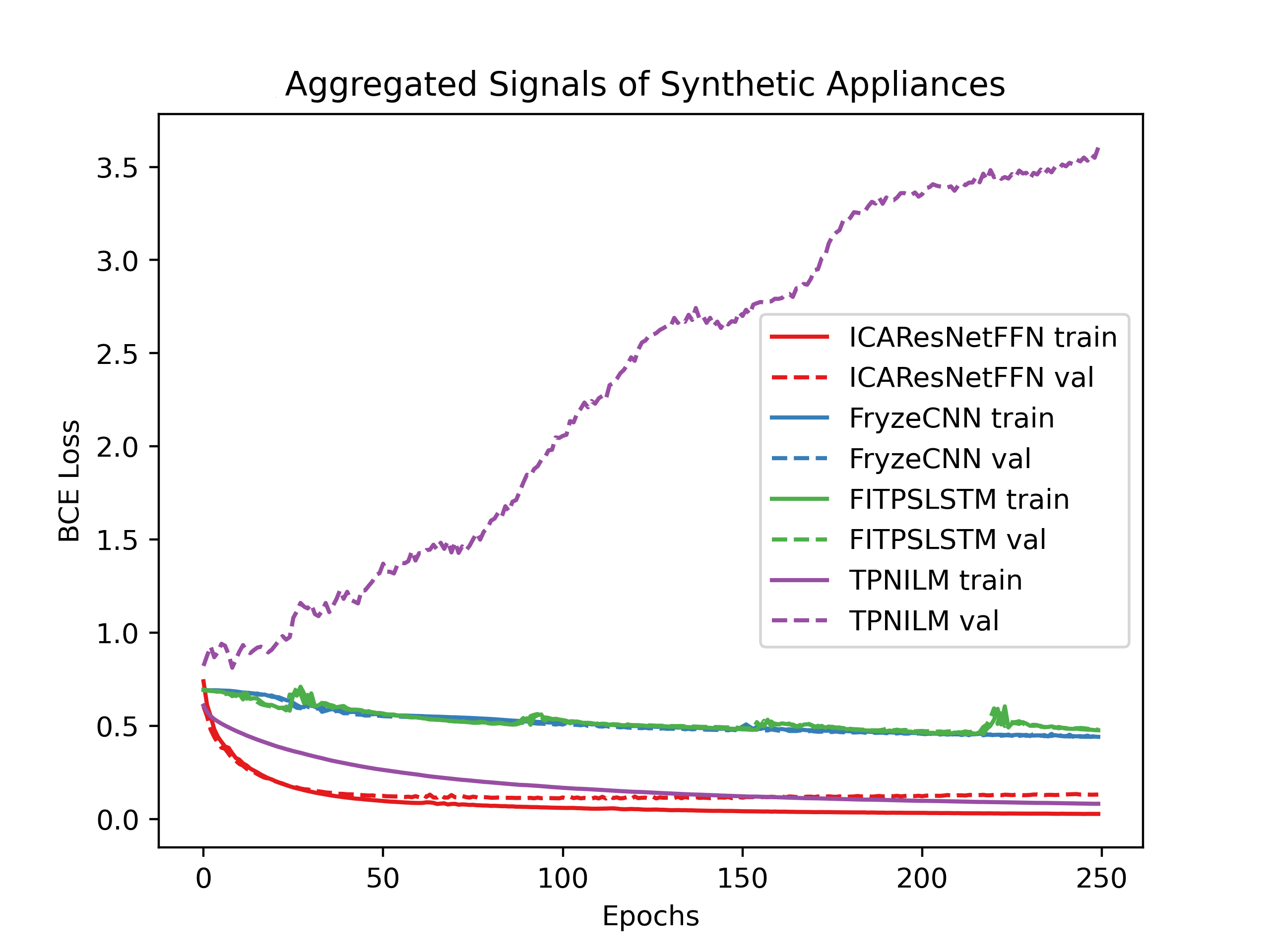}
    \caption{Binary cross entropy loss for four deep learning models.}
\label{fig:synth-loss}
\vskip -0.2in
\end{figure}

\begin{figure}
    \centering
    \includegraphics[width=\columnwidth]{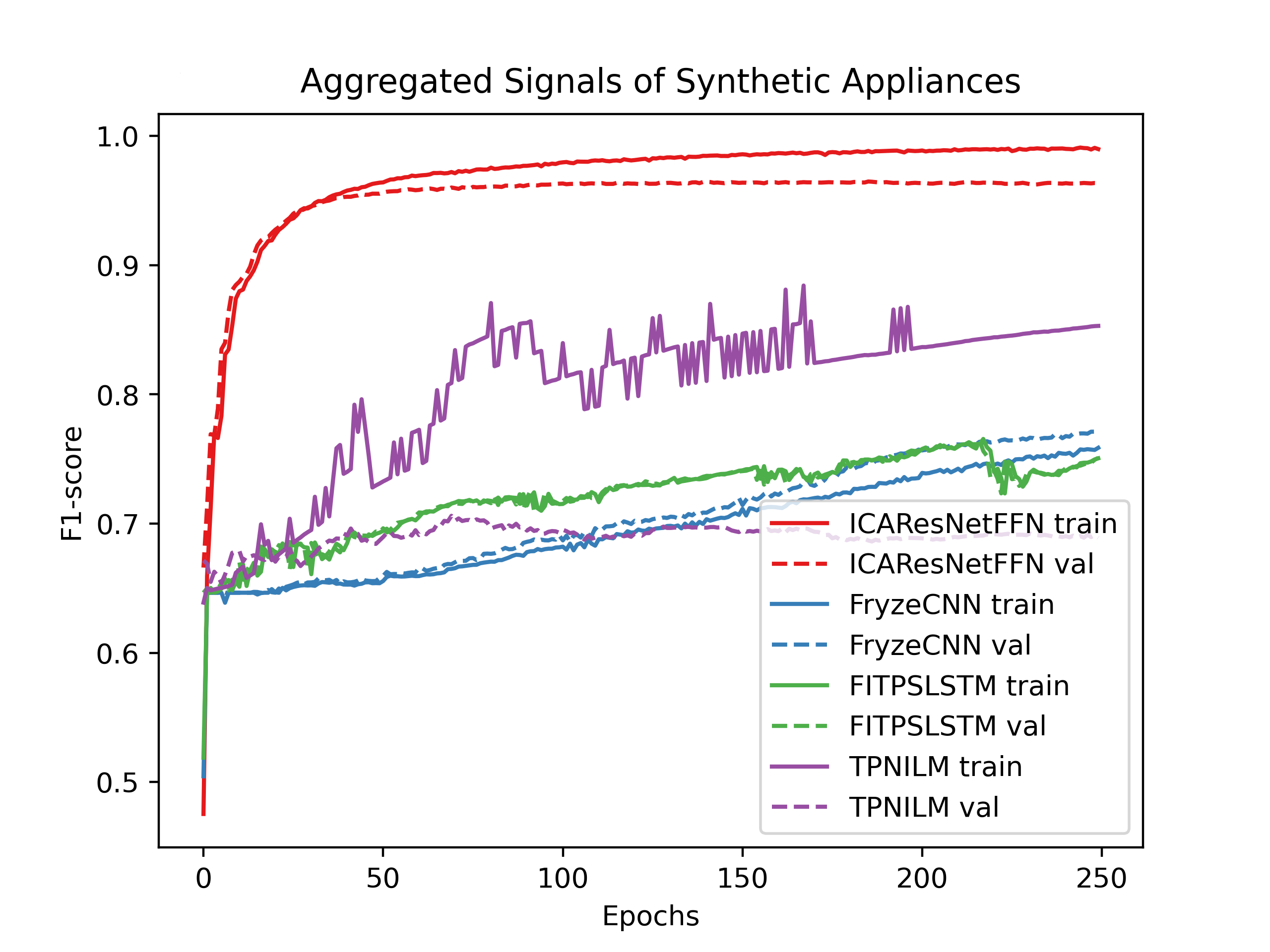}
    \caption{$F_1$-score (sample average) for four deep learning models.}
\label{fig:synth-f1}
\vskip -0.2in
\end{figure}

\begin{figure*}[t]
\centering
\includegraphics[width=\textwidth]{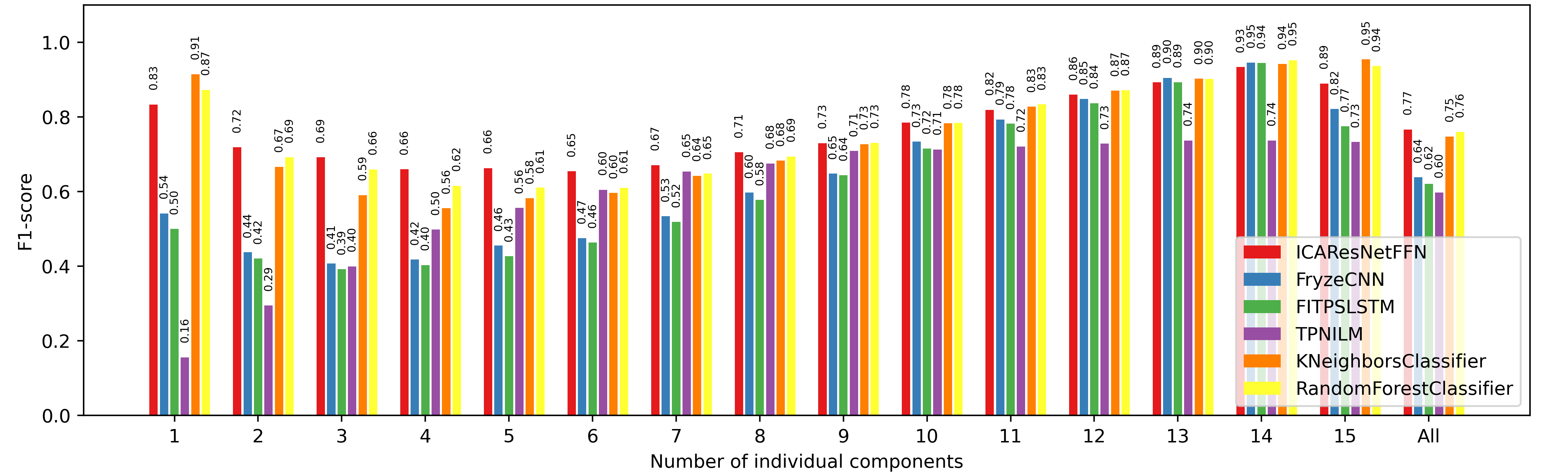}
\caption{Distribution of $F_1$-scores (samples averaging) across different number of simultaneously working appliances for four deep learning models and 2 classical machine learning models.}
\label{fig:synth-f1-bars}
\vskip -0.2in
\end{figure*}

Figures \ref{fig:synth-loss} and \ref{fig:synth-f1} depict the outcomes of training and validating four deep learning models using synthetic data. Our proposed model demonstrates the lowest validation loss and the highest $F_1$-score, exhibiting smoother convergence. Furthermore, Figure \ref{fig:synth-f1-bars} reveals that our model sustains a consistent $F_1$-score across varying numbers of concurrently operating appliances, whereas the other models exhibit a decline between two and ten individual components.

\subsection{Experiment 2: Real Appliances}
\begin{figure}[t]
    \centering
    \includegraphics[width=\columnwidth]{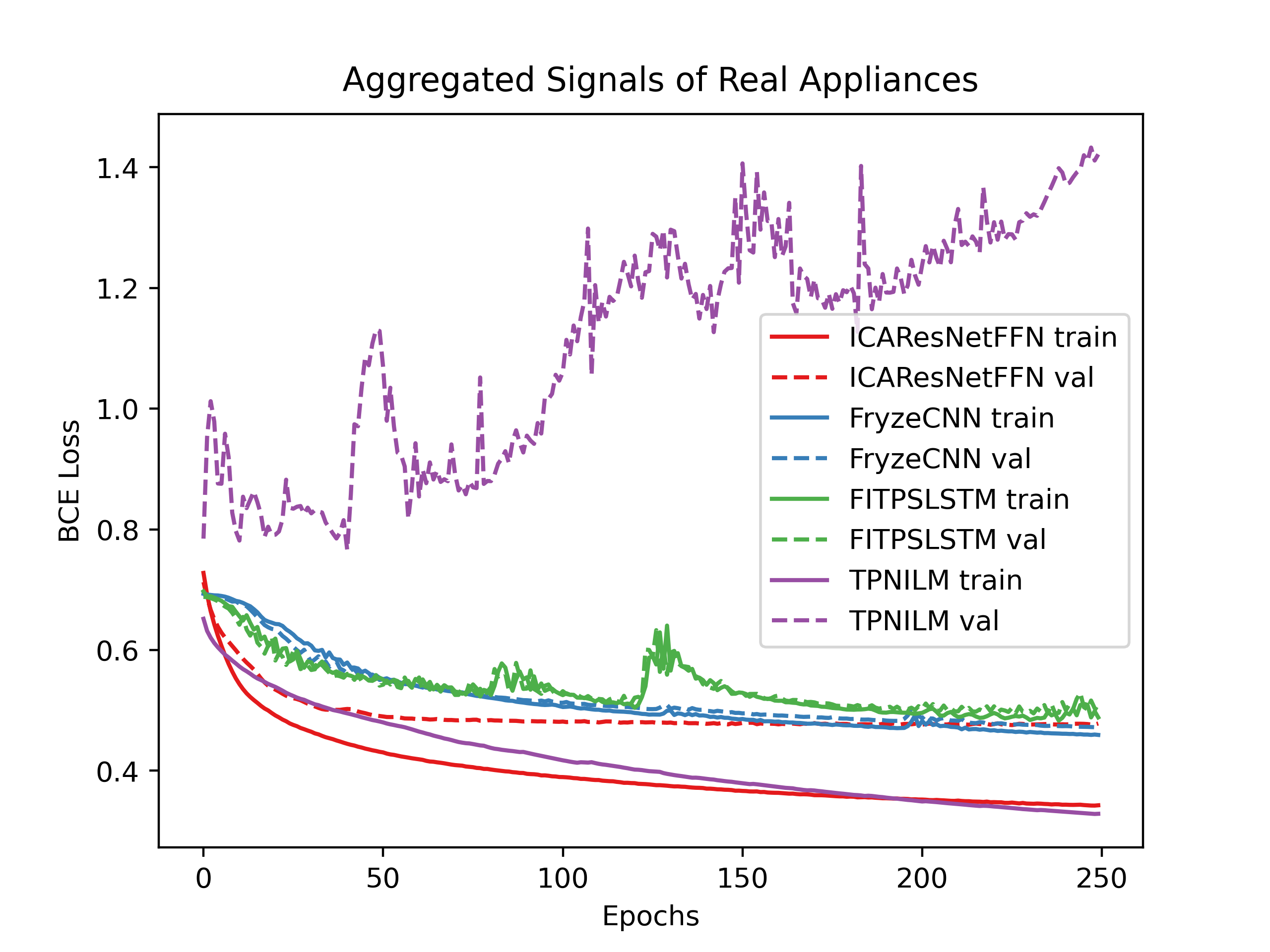}
    \caption{Binary cross entropy loss for four deep learning models.}
\label{fig:real-loss}
\vskip -0.2in
\end{figure}

\begin{figure}[t]
    \centering
    \includegraphics[width=\columnwidth]{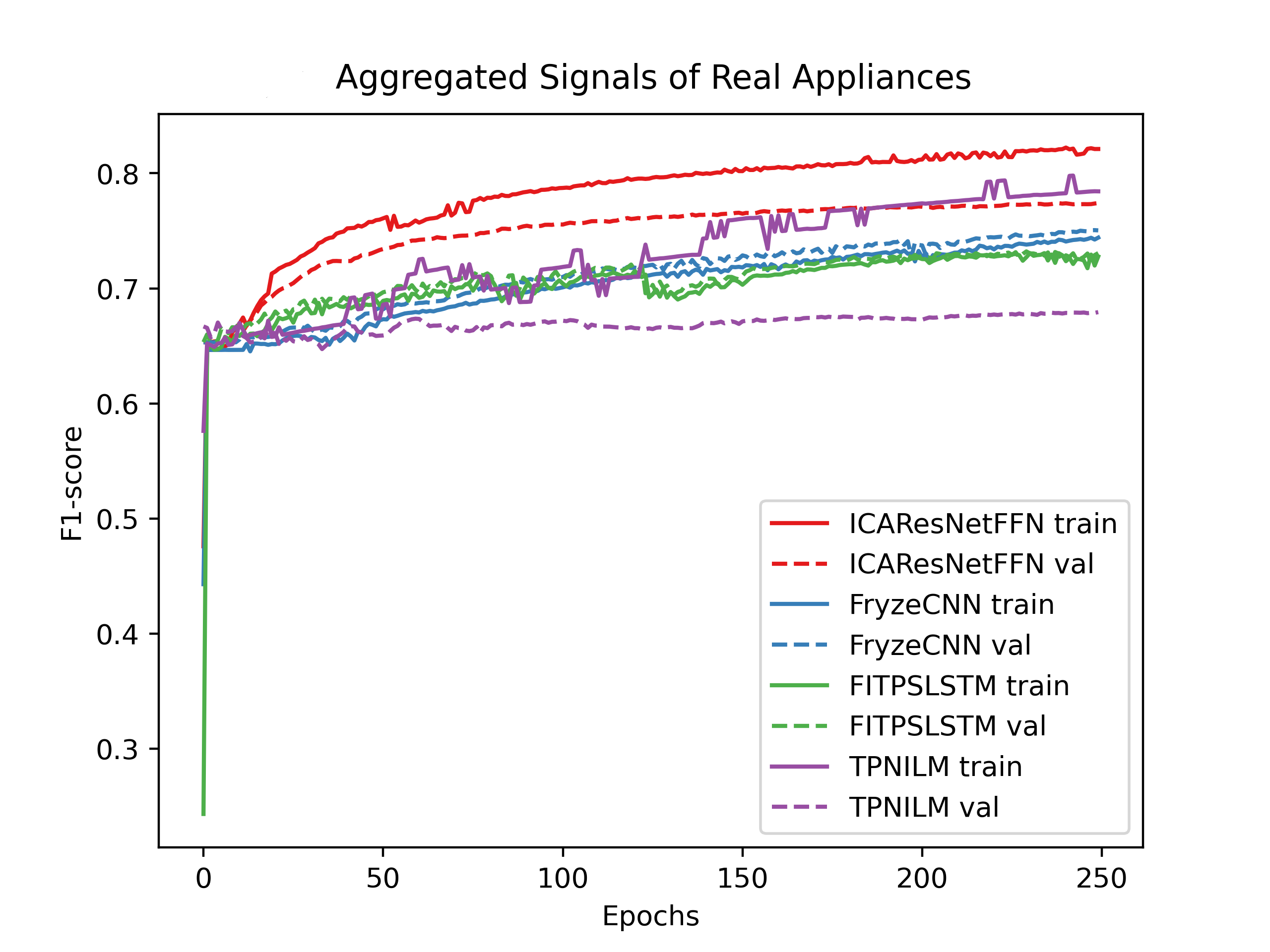}
    \caption{$F_1$-score (sample average) for four deep learning models.}
\label{fig:real-f1}
\vskip -0.2in
\end{figure}

\begin{figure*}[t]
\centering
\includegraphics[width=\textwidth]{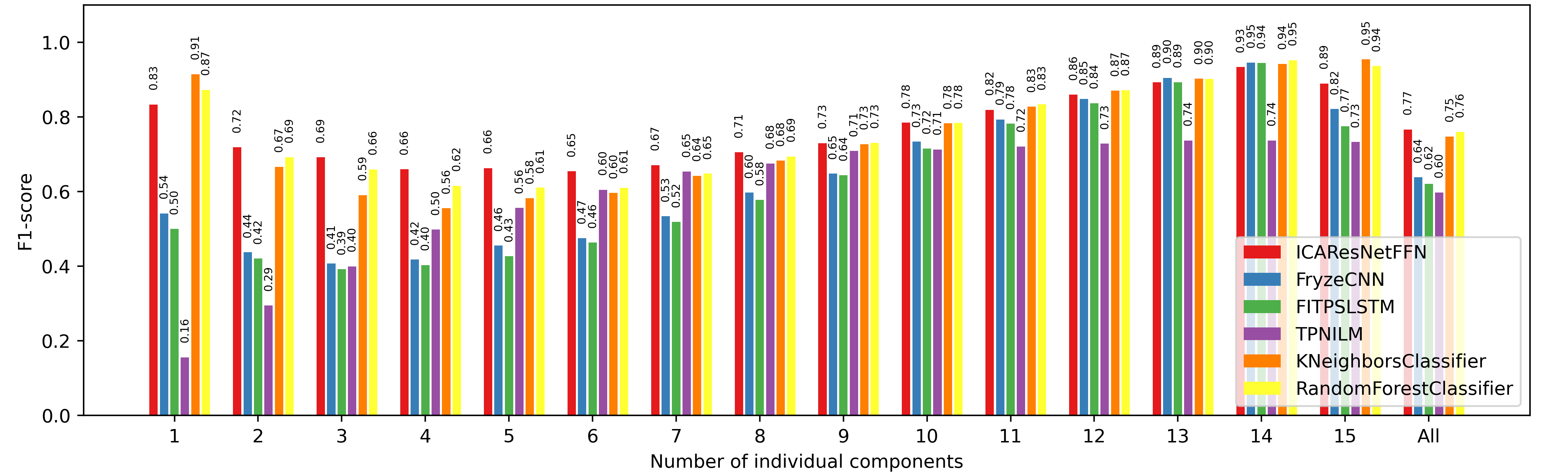}
\caption{Distribution of $F_1$-score (samples averaging) across different number of simultaneously working appliances for 4 deep learning models and 2 classical machine learning models.}
\label{fig:real-f1-bars}
\vskip -0.1in
\end{figure*}

Figures \ref{fig:real-loss} and \ref{fig:real-f1} display the binary cross-entropy loss and the $F_1$-score for training and validation of four deep learning models on real data, respectively. In this case also, ICA-ResNetFFN achieves the lowest validation loss and highest $F_1$-score, and exhibits smoother convergence compared to the other models. However, Figure \ref{fig:real-f1-bars} reveals that our model no longer sustains a uniform $F_1$-score across varying numbers of concurrently operating appliances. Nonetheless, it demonstrates a higher $F_1$-score where the other algorithms experience performance drops. The results are summarized in Table \ref{tab:my-table}, which compares the $F_1$-scores (sample averaging) of all models.

\begin{table*}[]
\centering
\caption{Average accuracy for all models}
\label{tab:my-table}
\begin{tabular}{lcccccc}
\toprule
\multicolumn{1}{c}{\textbf{Experiment}} &
  \multicolumn{1}{c}{\textit{ICA+ResNetFFN}} &
  \multicolumn{1}{c}{\textit{Fryze+CNN}} &
  \multicolumn{1}{c}{\textit{FIT-PS+LSTM}} &
  \multicolumn{1}{c}{\textit{Temporal Pooling NILM}} &
  \multicolumn{1}{c}{\textit{k-NN}} &
  \multicolumn{1}{c}{\textit{RandomForest}} \\
\midrule
\textbf{Real data}      & 0.77 & 0.64 & 0.62 & 0.6  & 0.75 & 0.76 \\ 
\textbf{Synthetic data} & 0.95 & 0.68 & 0.72 & 0.67 & 0.88 & 0.93 \\ 
\bottomrule
\end{tabular}
\end{table*}

\begin{figure*}
\centering
\includegraphics[width=\textwidth]{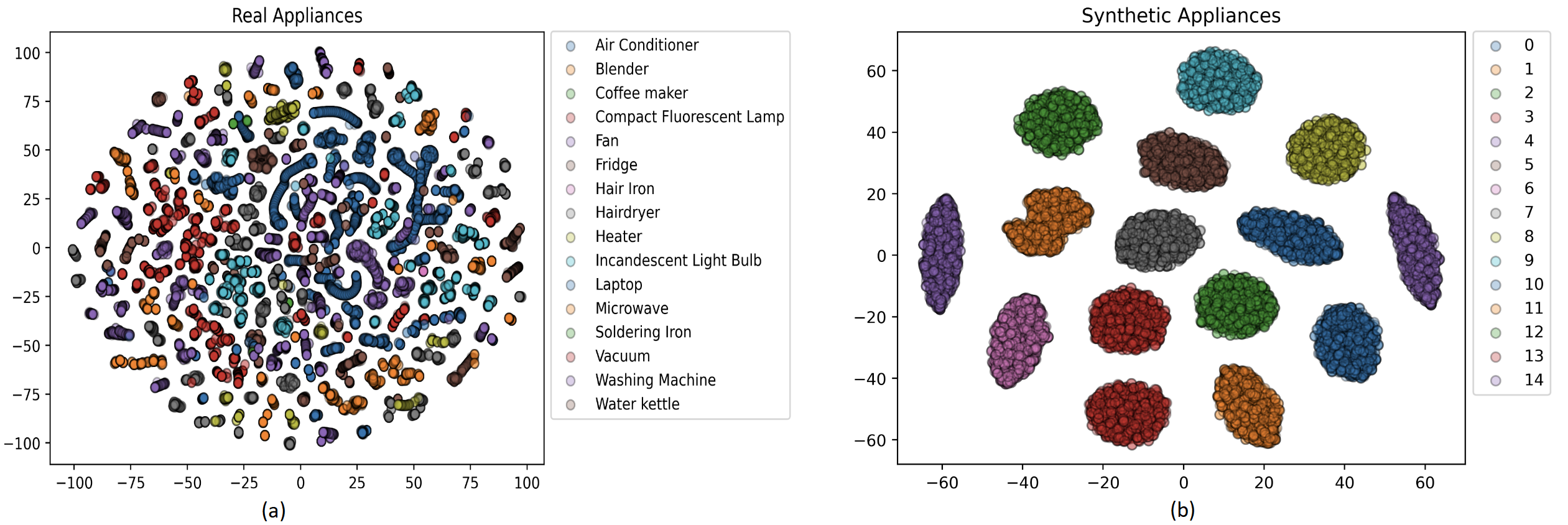}
\caption{t-SNE visualization of the feature space for real data (left), Synthetic data(right).}
\label{fig:mixed-tsne}
\vskip -0.2in
\end{figure*}

The t-SNE embeddings of real and synthetic appliance classes can justify the performance differences observed in the experiments. Figure \ref{fig:mixed-tsne}(a) reveals the complex structure of real appliance classes, with some classes exhibiting multiple data-point clusters or overlapping with others. 

This overlap occurs naturally due to appliances containing common electrical elements. For instance, washing machines have heating elements, motors, and water pumps, while heating elements are also the primary component of water kettles. Consequently, when operating simultaneously, these appliances may be misidentified as a single device, leading to reduced performance. Developing a disaggregation algorithm solely on real data may not accurately capture its generalization capability. Employing synthetic data with linearly separable classes, as shown in Figure \ref{fig:mixed-tsne}(b), can guide the development of an optimal architecture that can be subsequently applied to real data.

\section{Conclusion}\label{sec:conclusion}
In conclusion, the present work shows the importance of considering the underlying physics of the data when selecting an appropriate model to accurately capture the physical characteristics of the energy disaggregation problem.We employed Independent Component Analysis (ICA) as the feature extraction approach since it assumes that the signals are linearly mixed, which is consistent with Kirchhoff's circuit principles. A primary objective of this research was to evaluate the performance of the algorithm in scenarios with varying numbers of concurrent appliances. To ensure that our dataset was sufficiently diverse to accommodate all possible combinations of components, we deliberately curated a rich and comprehensive dataset. Our findings indicate that ICA exhibits exceptional performance when applied to large datasets with abundant training examples for any number of simultaneous appliances. Our model outperforms modern, existing baseline models. In addition, it better handles the problem of data imbalance, which was not properly solved until now. This study contributes valuable insights into the effective application of ICA for modeling and analyzing complex physical systems with multiple interacting components. It must be noted that this work is focused on classifying only three appliances; so the performance of the method needs to be checked consider a greater number of which is the object of future works.

\bibliography{references}
\bibliographystyle{IEEEtran}

\end{document}